\documentclass[%showpacs, 
superscriptaddress,twocolumn, preprintnumbers, nofootinbib,amsmath,amssymb]{revtex4}
\usepackage{graphicx}% Include figure files
\usepackage{dcolumn}% Align table columns on decimal point
\usepackage{bm}% bold math
\usepackage{soul}
\usepackage{xcolor}

\newcommand\ee{\end{equation}}
\newcommand\be{\begin{equation}}
\newcommand\eea{\end{eqnarray}}
\newcommand\bea{\begin{eqnarray}}

\newcommand\mpl{M_{\rm pl}}

\definecolor{DarkBlue}{rgb}{0.15,0.15,0.85}
\usepackage[linktocpage=true]{hyperref}
\hypersetup{
colorlinks=true,
citecolor=DarkBlue,
linkcolor=DarkBlue,
urlcolor=DarkBlue,
}

\begin{document}

%\preprint{}

\title{$\phi^2$ Inflation at its Endpoint}% Force line breaks with \\

\author{Paolo Creminelli}
\affiliation{Abdus Salam International Centre for Theoretical Physics,\\ Strada Costiera 11, 34151, Trieste, Italy}
\affiliation{Institute for Advanced Study, Princeton, New Jersey 08540, USA}
\author{Diana L\'opez Nacir}
\affiliation{Abdus Salam International Centre for Theoretical Physics,\\ Strada Costiera 11, 34151, Trieste, Italy}
\affiliation{Departamento de F\'isica and IFIBA, FCEyN UBA, Facultad de Ciencias Exactas y Naturales, Ciudad Universitaria, Pabell\'on I, 1428 Buenos Aires, Argentina}
\author{Marko Simonovi\'c}
\affiliation{SISSA, via Bonomea 265, 34136, Trieste, Italy}
\affiliation{Istituto Nazionale di Fisica Nucleare, Sezione di Trieste, I-34136, Trieste, Italy}
\author{Gabriele Trevisan}
\affiliation{SISSA, via Bonomea 265, 34136, Trieste, Italy}
\affiliation{Istituto Nazionale di Fisica Nucleare, Sezione di Trieste, I-34136, Trieste, Italy}
\author{Matias Zaldarriaga}
\affiliation{Institute for Advanced Study, Princeton, New Jersey 08540, USA}


\begin{abstract}
In the simplest inflationary model $V=\frac12 m^2\phi^2$, we provide a prediction accurate up to $1\%$ for the spectral index $n_s$ and the tensor-to-scalar ratio $r$ assuming instantaneous reheating and a standard thermal history: $n_s = 0.9668\pm0.0003$ and $r=0.131\pm 0.001$. This represents the simplest and most informative point in the $(n_s,r)$ plane. The result is independent of the details of reheating (or preheating) provided the conversion to radiation is sufficiently fast. A slower reheating or a modified post-inflationary evolution push towards smaller $n_s$ (and larger $r$), so that our prediction corresponds to the maximum $n_s$ (and minimum $r$) for the quadratic potential.  We also derive similar results for a general $V \propto \phi^p$ potential.

\end{abstract}

%\pacs{\dots}% PACS, the Physics and Astronomy
\maketitle
%\vspace*{0.3cm}

%%%%%%%%%%%%%%%%%%%%%%%%%%%%%%%%%%%%%%%%%%%%%%%%%%%%%%%%%
%%%%%%%%%%%%%%%%%%%%%%%%%%%%%%%%%%%%%%%%%%%%%%%%%%%%%%%%%
{\em Introduction.}---Recently,   the  BICEP2 collaboration \cite{Ade:2014xna}  has detected  B-modes in the polarization of the cosmic microwave background (CMB) at large angular scales. This signal has been  interpreted both as a detection of primordial gravitational waves and as microwave emission by polarized dust \cite{Mortonson:2014bja}.   Although we have to wait for the dust to settle, this detection motivates to explore the  consequences of a large tensor-to-scalar ratio $r$.  
One relevant point is that if $r \sim \mathcal{O}(0.1)$,
 it will be possible in future observations to ultimately measure $r$ with a precision of $1\%$ \cite{Dodelson:2014exa}.
This precision requires both to go to second order in slow-roll and to know the number of e-folds $N$ up to $\Delta N \sim 0.5$ \cite{Creminelli:2014oaa}. Thus, one has to specify the details of reheating (for a recent study see \cite{Dai:2014jja}) and include subleading corrections in the predictions of $r$ and $n_s$, which are usually neglected.

In this paper we focus on the inflationary model $V=\frac12m^2\phi^2$, both because it is simple and because it is the only case in which predictions can be made without further assumptions about the behaviour of the potential between the inflationary region and the final minimum.    
We provide improved formulas for $n_s$ and $r$ that are correct up to $1 \%$ relative errors, in the limit of fast reheating (the details turn out to be irrelevant as long as the energy is transferred to radiation fast enough) and with a standard expansion history after inflation. 
Deviations from this scenario, such as a slower reheating, entropy injections due to phase transitions, higher number of degrees of freedom at reheating, additional periods of matter dominance or inflation, move the predictions in the direction of higher $r$ and lower $n_s$. In this sense, the  values for $n_s$ and $r$ we provide are the ``endpoint" on the line in the $(n_s,r)$ plot for $V=\frac12m^2\phi^2$, corresponding to the largest possible $N$. This point deserves special attention, because in some sense it is the most informative place in the $(n_s,r)$ plane: if  data eventually converge there, it will be possible to have strong bounds on any deviation from this minimal scenario. 

\vspace{.3cm}
%%%%%%%%%%%%%%%%%%%%%%%%%%%%%%%%%%%%%%%%%%%%%%%%%%%%%%%%%
%%%%%%%%%%%%%%%%%%%%%%%%%%%%%%%%%%%%%%%%%%%%%%%%%%%%%%%%%
{\em Predictions for instantaneous reheating.}---In order to determine the point in the $(n_s,r)$ plane that corresponds to the limit of instantaneous reheating, we will start the calculation in the usual way and refine some of the steps. Take a mode with comoving momentum $k_*$, which crosses the horizon during inflation when the scale factor is $a_*$, $k_*=a_*H_*$. We want to compare the  wavelength of this mode with the size of the horizon today
\be
\label{N_start}
\frac{k_*}{a_0H_0} = \frac{a_*}{a_{\mathrm{end}}} \frac{a_\mathrm{end}}{a_{\mathrm{rh}}} \frac{a_{\mathrm{rh}}}{a_0} \frac{H_*}{H_0},
\ee
where $a_{\mathrm{end}}$ is the scale factor at the end of inflation and $a_{\mathrm{rh}}$ the scale factor when radiation starts to dominate. Of course this splitting in various phases is somewhat arbitrary and one expects to introduce errors of order $\Delta N \sim 1$. However, we will show later, using numerical solutions, that our analytical calculations are accurate up to $1\%$.
In our analytical calculation, we will assume instantaneous reheating: $a_{\mathrm{end}} = a_{\mathrm{rh}}$. Under these assumptions Eq.~\eqref{N_start} becomes
\be
\label{N_step2}
\frac{k_*}{a_0H_0} = e^{-N} \frac{a_\mathrm{rh}}{a_0} \frac{H_*}{H_0},
\ee
where $N$ is the number of e-folds from the moment the mode $k_*$ crosses the horizon until the end of inflation. Let us calculate the different terms  on the rhs of Eq.~\eqref{N_step2}.

The number of e-folds is given by
\be\label{efolds}
N = \int_{t_*}^{t_\mathrm{end}} H \mathrm d t =  \int_{\phi_\mathrm{end}}^{\phi_*} \frac 1{\sqrt{2\epsilon_\mathrm H }} \frac{\mathrm d \phi}{\mpl}.
\ee
The slow-roll parameter $\epsilon_\mathrm{H}$ is related to the derivatives of the potential as
\be
\epsilon_\mathrm H \equiv - \frac{\dot H}{H^2} = \epsilon_\mathrm V \left( 1- \frac 43 \epsilon_\mathrm V + \frac 23 \eta_\mathrm V \right),
\ee
where $\epsilon_\mathrm V$ and $\eta_\mathrm V$ are defined as
\be
\epsilon_\mathrm V \equiv \frac{1}{2}\mpl^2 \left( \frac{V'}{V} \right)^2, \quad \quad \eta_\mathrm V \equiv \mpl^2 \frac{V''}{V} .
\ee
For $V=\frac12 m^2\phi^2$, we have $\epsilon_\mathrm V=\eta_\mathrm V = 2\mpl^2/\phi^2$, and the number of e-folds from Eq.~(\ref{efolds}) is given by
\be
\label{Nefolds}
N = \frac{1}{4} \frac{\phi_*^2 - \phi_\mathrm{end}^2}{\mpl^2} + \frac 13 \log \frac{\phi_*}{\phi_\mathrm{end}}+\cdots.
\ee
The first term is the standard result, while the second is the next to leading order in the slow-roll expansion. Notice that typically, given that $\phi_*\gg\phi_\mathrm{end}$, both $\phi_\mathrm{end}^2$ and the logarithmic correction  are dropped in the usual calculation, but we are going to keep them here. Of course, the formula above must break down towards the end of inflation because higher slow-roll corrections (encoded in $+\cdots$) become important. Naively this can change the result for the number of e-folds by order one. However, as we will show later, Eq.~\eqref{Nefolds} appears to be an excellent approximation to the numerical solutions.

The first fraction on the rhs of Eq.~\eqref{N_step2} can be evaluated using entropy conservation. Here we are assuming that none of the processes in the early universe lead to an entropy injection. This is a good approximation for the known phase transitions (electroweak and QCD).
If entropy is conserved, then
\be
g_{*}^\mathrm{rh} a_\mathrm{rh}^3 T_\mathrm{rh}^3 = a_0^3T_\gamma^3 \left( 2  + \frac{4}{11} g_{*}^\nu \right),  
\ee
where $g_{*}^\mathrm{rh}$ is the number of degrees of freedom at the end of inflation, $g_{*}^\nu$ the number of degrees of freedom of neutrinos, $T_\gamma$ is the temperature of the CMB photons today, and we have set the temperature of neutrinos to $T_\nu^3= \frac{4}{11} T_\gamma^3$. For the $3$ neutrino species, $g_{*}^\nu=21/4$, and the first ratio on the rhs of Eq.~\eqref{N_step2} becomes
\be
\frac{a_\mathrm{rh}}{a_0} = \frac{T_\gamma}{T_\mathrm{rh}} \left( \frac{43}{11g_{*}^\mathrm{rh}} \right)^{1/3}.
\ee

To calculate the temperature at the beginning of radiation dominance, we will assume that inflation ends when $\ddot a=0$. Although this definition is arbitrary, one can check that analytical results with different choices of the point where inflation ends (some popular choices are $\epsilon_\mathrm V=1$ or $\phi_\mathrm{end}=1\mpl$) give the same predictions within the precision we are working at. Assuming that $\dot\phi$ has the attractor value $\dot\phi = - \sqrt {2/3}\,m\mpl$, from the relation 
\be
\dot H = \frac{\ddot a}{a} - H^2 = -\frac 12 \frac{\dot\phi^2}{\mpl^2},
\ee
one finds that inflation ends at $\phi_\mathrm{end}=\sqrt{4/3}\,\mpl$. The energy density at the end of inflation is
\be
\rho_\mathrm{end} = \frac 12 m^2 \phi_\mathrm{end}^2 + \frac 12 \dot\phi^2 = m^2\mpl^2.
\ee
In our toy model this energy density is instantaneously converted into radiation with temperature
\be
T_\mathrm{rh} = \left( \frac{30 m^2\mpl^2}{\pi^2 g_{*}^\mathrm{rh}} \right)^{ 1/4}.
\ee
This fixes the ratio $a_\mathrm{rh}/a_0$.

The last fraction on the rhs of Eq.~\eqref{N_step2} depends on $H_*$
\be
H_* =\frac{m}{3} \sqrt{1+\frac 32 \frac{\phi_*^2}{\mpl^2}} \approx \frac{m\phi_*}{\sqrt 6 \mpl}\;.
\ee
The mass of the inflaton is determined from the normalization of the power spectrum 
\be
\label{noramlization}
\Delta_\zeta^2 = \frac{k^3}{2\pi^2} P_\zeta = \frac{1}{96\pi^2} \frac{\phi_*^4}{\mpl^4}\frac{m^2}{\mpl^2}.
\ee
Replacing all previous results in Eq.~\eqref{N_step2}, we obtain a relation between $\phi_*$ and $k_*$
\begin{align}
\label{all_parameters}
& \log \frac{k_*}{a_0H_0}= \frac{1}{4} \frac{\phi_\mathrm{end}^2 - \phi_*^2}{\mpl^2} - \frac 13 \log \frac{\phi_*}{\phi_\mathrm{end}} + \log\frac{T_\gamma}{H_0} \nonumber \\
& \quad -\frac{1}{12} \log g_{*}^\mathrm{rh}+ \frac{1}{4}\log \Delta_\zeta^2 + \log\frac{4\pi(43/11)^{1/3}}{2880^{1/4}}.
\end{align}

Setting $g_{*}^\mathrm{rh}= 106.75$, $H_0=1.5\times 10^{-42}\;\mathrm{GeV}$, $T_\gamma=0.235\times 10^{-12} \; \mathrm{GeV}$ and $\Delta_\zeta^2=2.19\times 10^{-9}$ $(\footnote{Strictly speaking, this value of $\Delta_\zeta^2$ is given for $k=0.05\;\mathrm{Mpc}^{-1}$. Although $k_*$ can be different from this scale, $\Delta_\zeta$ appears only inside logarithms and the error one makes is much smaller than the precision we want to achieve.})$ \cite{Ade:2013zuv} we get
\be
\label{final_phi}
\log \frac{k_*}{a_0H_0} = -\frac 14 \frac{\phi_*^2}{\mpl^2} - \frac 13 \log \frac{\phi_*}{\mpl} + 63.3.
\ee
The logarithmic contribution in this equation is very small and a very good approximate solution is 
\be
\frac{\phi_*}{\mpl} = \alpha -\frac{2}{3\alpha}\log \alpha\;, \quad \alpha = 2 \left( 63.3-\log\frac{k_*}{a_0H_0} \right)^{1/2}.\nonumber
\ee
For the pivot scale $k_*=0.002 \; \mathrm{Mpc}^{-1}$, the numerical value for $\alpha$ is $\alpha=15.6$.
Once $\phi_*$ is known, it is an easy exercise to calculate the tensor-to-scalar ratio and the spectral index. For a quadratic potential  $\epsilon_\mathrm V=\eta_\mathrm V$, and at second order in slow-roll \cite{Stewart:1993bc}

\begin{align}
r &= 16 \epsilon_\mathrm{V} \left( 1-\frac 23 \epsilon_\mathrm{V} + 2 C \epsilon_\mathrm{V} \right), \nonumber \\
n_s &= 1-4 \epsilon_\mathrm{V} - \epsilon_\mathrm{V}^2 \left( 8C + \frac{14}{3} \right),
\end{align}
where $C=-2+\log 2 +\gamma$ and $\gamma=0.75521\ldots$ is Euler-Mascheroni constant. 
The prediction for the endpoint of the quadratic inflationary potential in the limit of instantaneous reheating is
\be
\label{prediction}
n_s = 0.9668\pm0.0003, \quad \mathrm{and} \quad r=0.131\pm 0.001,
\ee
for $k_*=0.002 \; \mathrm{Mpc}^{-1}$. This point corresponds to the number of e-folds $N=60.7\pm0.5$. 
 
The point in the $(n_s,r)$ plane defined by these two values corresponds to the maximum value of $N$ for the quadratic potential: {\em changes with respect to this minimal picture decrease $N$.}  A slower reheating gives a lower $N$. The same happens if we increase the number of relativistic degrees of freedom at reheating. The dependence of $N$ on $g_*$ is anyway very mild, $N \sim -\frac{1}{12} \log{g_*}$, see Eq.~\eqref{all_parameters}, and one should vary $g_*$ by orders of magnitude to get a relevant effect. Another effect that pushes towards a smaller value of $N$ is entropy injection during the thermal history of the Universe. For example this happens with a first-order phase transition or when a massive particle goes out of thermal equilibrium before decaying. Notice also that a ``standard'' equation of state $w\leq 1/3$ before thermalization pushes again towards smaller values of $N$. The only way to get to larger values is to have $w>1/3$ after inflation, which is a somewhat exotic possibility\footnote{For example one can have a period of kinetic domination ($w=1$) after the end of the slow-roll regime: after that the inflaton can get trapped in a minimum and reheat the Universe. Even if we allow for values of $w$ larger than $1/3$, one can still get an absolute bound on the value of $N$. This is achieved in the extreme case where $w\gg1$ from the end of inflation until BBN: the maximum value is $N\approx 100$ and this corresponds to $n_s=0.98$ and $r=0.08$.}.   

\vspace{.3cm}
%%%%%%%%%%%%%%%%%%%%%%%%%%%%%%%%%%%%%%%%%%%%%%%%%%%%%%%%%
%%%%%%%%%%%%%%%%%%%%%%%%%%%%%%%%%%%%%%%%%%%%%%%%%%%%%%%%%
{\em Numerical checks.}---The predictions for $n_s$ and $r$ are derived under a number of assumptions. Given the level of  precision  we are working at, one might be worried that the corrections to the analytical calculation are large enough to spoil the final result. There are several possible sources of errors and we discuss them in this section.

The first one is the issue of matching  different phases of evolution before thermalization. One would expect a sharp matching between different phases to give an error of order $\Delta N \sim1$, and this would be relevant for the precision we want to achieve. The other problem is that it is not obvious how our final result depends on the details of reheating or preheating. To address these two questions we numerically solved a toy model that describes the evolution of the inflaton field and of radiation with  energy density $\rho_r$
\begin{align}
\label{toy}
& \ddot \phi + 3H\dot \phi +\Gamma \dot\phi +m^2 \phi =0 , \nonumber \\
& \dot\rho_r + 4H\rho_r - \Gamma \dot\phi^2 = 0  , \nonumber \\
& 3\mpl^2 H^2 = \frac 12 \left( m^2\phi^2 + \dot\phi^2 \right) + \rho_r .
\end{align}
$\Gamma$ is a constant that characterises the efficiency of the transfer of energy from the inflaton field to radiation. Of course, this is not a realistic model, especially when preheating effects are relevant, but it will be sufficient to show that the details of the transition are not relevant, provided it is fast enough.

\begin{figure}[!!!h]
\begin{center}
\includegraphics[width=0.485 \textwidth]{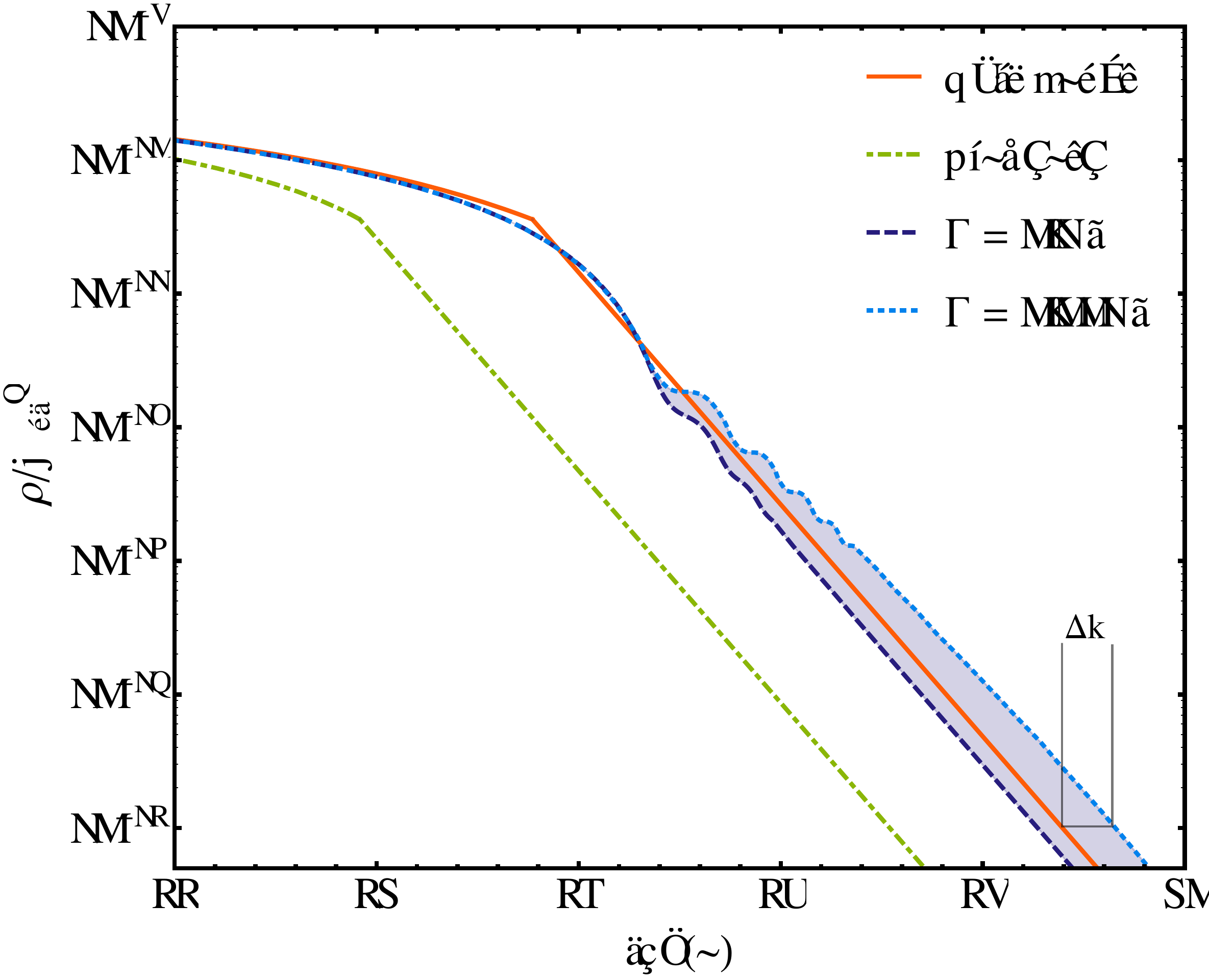}
\end{center}
\caption{\small {\em Energy density as a function of $\log (a)$. Green dot-dashed line: the standard analytical result. Red solid line: analytical result using Eq.~\eqref{Nefolds}. The grey region corresponds to numerical solutions between $\Gamma=m/10$ and $\Gamma=m/1000$. The improved analytical formula is within $0.3$ e-folds from the numerical solutions for this range of $\Gamma$.}}
\label{fig:rhologa}
\end{figure}

In Fig.~\ref{fig:rhologa} we show how the total energy density depends on $\log (a)$. The initial conditions for the numerical integration are $\phi_*=15\;\mpl$, $\dot\phi = - \sqrt {2/3}\,m\mpl$ (i.e.~on the attractor solution) and $\rho_r(a_*)=0$. We take $a_*=1$, $m=6\times 10^{-6}\mpl$, and turn on $\Gamma$ when $\ddot a =0$ (we are going to comment on this choice later). Fig.~\ref{fig:rhologa} shows the numerical results for a range of $\Gamma$ from $\Gamma=m/1000$ (cyan dotted line) to $\Gamma=m/10$ (blue dashed line), while the analytical solution, using Eq.~\eqref{Nefolds}, is represented by the red solid line. For this range of parameters analytical and numerical results agree within $\Delta N = 0.3$, once radiation dominance is reached. This agreement is better than what one would naively expect and justify the use of the analytic formula. Therefore, as long as $\Gamma\gtrsim m/1000$, the transition to radiation can be considered instantaneous in order to make predictions with $1\%$ accuracy. We also plot the result using the standard analytical formula Eq.~\eqref{Nefolds} without the logarithmic term (green dot-dashed line). The difference compared to numerical solutions is $\Delta N \approx 1$, and it is not accurate enough for the precision we are working at. 

\begin{figure}[!!!h]
\begin{center}
\includegraphics[width=0.485 \textwidth]{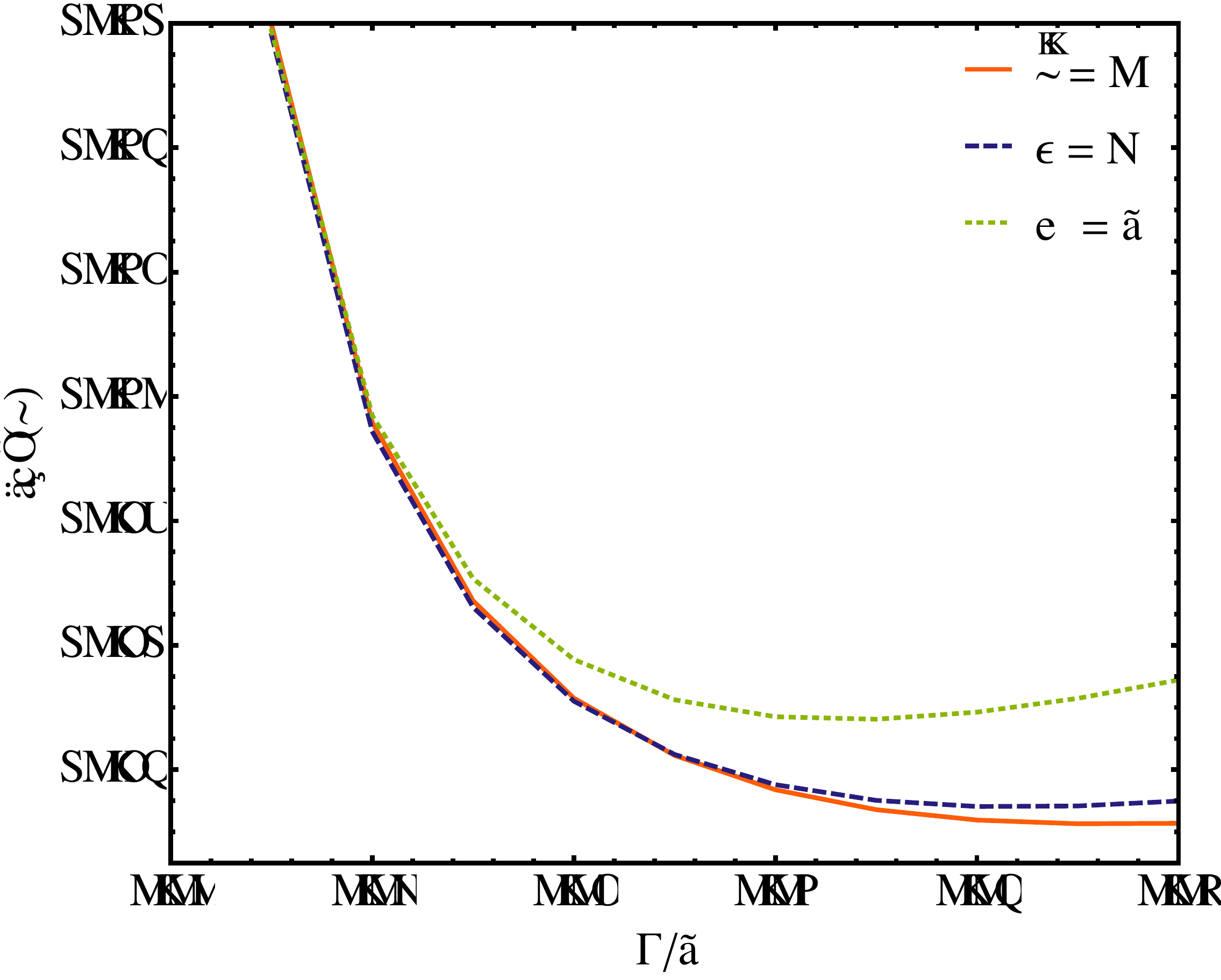}
\end{center}
\caption{\small {\em Dependence of the expansion of the universe $\log (a)$ on $\Gamma$ for fixed initial conditions and fixed final energy density (in radiation dominance). Different curves are obtained for different choices of time at which the inflaton starts to decay.}}
\label{fig:gammadep}
\end{figure}

In Fig.~\ref{fig:gammadep} we plot  the expansion $\log (a)$ required to reach a fixed final energy density (in radiation dominance) as a function of $\Gamma$, for fixed initial conditions. We note that it flattens out for sufficiently large values of $\Gamma$,  indicating that the number of e-folds $N$ becomes insensitive to $\Gamma$ in that regime\footnote{The mild raise at large $\Gamma$ of the curve corresponding to $H=m$ is due to the friction induced on the inflaton by the production of radiation.}. To assess the relevance of our choice of turning on $\Gamma$ when $\ddot a=0$,  in Fig.~\ref{fig:gammadep} we also plot the curves corresponding to the  conditions $\epsilon=1$ and $H=m$.  The  alternative conditions give a negligible difference in the amount of expansion. For small values of $\Gamma$ the curve steepens and approximately approaches the analytical expectation $a \propto \Gamma^{-1/6}$. 

\begin{figure}[!!!h]
\begin{center}
\includegraphics[width=0.485 \textwidth]{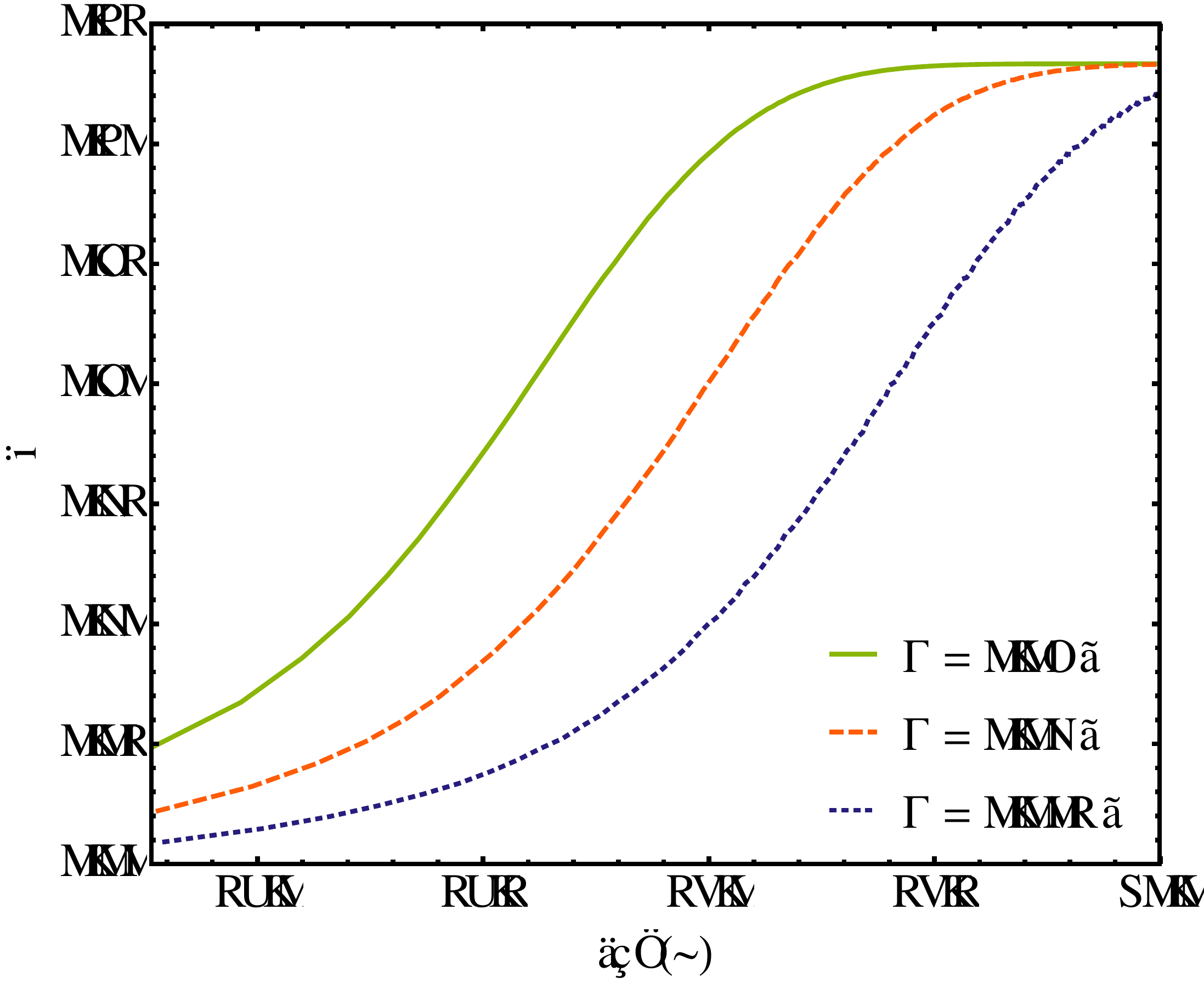}
\end{center}
\caption{\small {\em Evolution of the equation of state (averaged over a period of oscillation) as a function of $\log (a)$ for different values of $\Gamma$.}}
\label{fig:wloga}
\end{figure}

One may wonder whether our results apply when the energy transfer to radiation occurs through a preheating stage. In this case the energy is efficiently converted to light particles and the equation of state quickly approaches $w=1/3$, even if the complete termalization of the system will occur much later \cite{Podolsky:2005bw}. To verify that, within the required accuracy, the phenomenological model of Eq.~\eqref{toy} gives results which are quite similar to what happens with preheating, we plot in Fig.~\ref{fig:wloga} the evolution of the equation of state $w$ for several values of $\Gamma$ (for which our prediction \eqref{prediction} applies, see Fig.~\ref{fig:rhologa}). This can be compared with the analogous results of \cite{Podolsky:2005bw,Dufaux:2006ee}: in both cases the equation of state reaches $w \sim 1/3$ in approximately $3$ e-folds.  Therefore preheating models, for a wide range of couplings, give a sufficiently fast reheating and lead to our result \eqref{prediction}.   

It is worth stressing that preheating makes an efficient conversion to radiation rather natural and compatible with the approximate shift symmetry that keeps flat the inflaton potential. Indeed a coupling $g^2 \phi^2 \chi^2$ induces a quick conversion to $\chi$ particle for $g^2 \gtrsim 10^{-10}$. For these small couplings one can safely neglect the radiative corrections to the quadratic potential \cite{Podolsky:2005bw}.

\vspace{.3cm}

{\em  Other monomial potentials.}---Everything we have said so far can be generalized to other monomial potentials $V=M^{4-p}\phi^p$ (we focus on $p \leq 2$, given the present experimental bounds). However, unlike the case of a quadratic potential, we expect that a generic monomial cannot be extrapolated to the origin. In particular, we expect the potential to steepen approaching the minimum: this steepening {\em increases} $N$ and can thus push the predictions beyond the would-be endpoint of the exact $\phi^p$ potential. In the following we assume that the modification of the potential happens at $\phi \ll \mpl$, so that it does not affect the predictions. Otherwise one should have control of the shape of the potential until the final minimum \cite{Silverstein:2008sg,McAllister:2008hb,McAllister:2014mpa}.

Eq.~\eqref{all_parameters} generalizes to 
\begin{align}
\label{result_p}
& \log \frac{k_*}{a_0H_0}= \frac{1}{2p} \frac{\phi_\mathrm{end}^2 - \phi_*^2}{\mpl^2} + \frac{3p-10}{12}\log \frac{\phi_*}{\mpl}  \nonumber \\
& \quad + \frac 13 \log \frac{\phi_\mathrm{end}}{\mpl} +\log\frac{T_\gamma}{H_0}-\frac{1}{12} \log g_{*}^\mathrm{rh} + \frac{1}{4}\log \Delta_\zeta^2\nonumber \\
& \quad     + \log \frac{2\pi(43/11)^{1/3} \sqrt{p}}{540^{1/4} \left( p/\sqrt{3}\right)^{p/4}},
\end{align}
with $\phi_\mathrm{end}=p/\sqrt{3}\mpl$. 
For instance, in the cases of $V\propto \phi$ and $V\propto \phi^{2/3}$, motivated by string-theory monodromy \cite{Silverstein:2008sg,McAllister:2008hb,McAllister:2014mpa}, the predictions for the endpoint are
\begin{align}
p=1&: \quad n_s = 0.9749, \quad r=0.0664, \quad N=60.0, \nonumber \\
p=2/3&: \quad n_s = 0.9776, \quad r=0.0446, \quad N=59.7.\nonumber
\end{align}
Independently of $p$, $n_s-1$ and $r$ have relative errors $\lesssim 1 \%$.

\vspace{.3cm}

{\em Conclusions.}---For a generic potential it is of limited interest to consider second order slow-roll corrections or changes in the number of e-folds $\Delta N \simeq 1$, because the effects are degenerate with small changes in the shape of the potential. However, this is relevant for the ``simplest" inflationary model, i.e.~$V \propto \phi^2$ with fast reheating, and one can make predictions with $1\%$ accuracy. This represents in some sense the most informative point in the $(n_s,r)$ plane: if future data will be compatible with it, we will be quite confident about the inflationary potential, the reheating process and the following thermal history. 

\vspace{.3cm}

\noindent
{\em Acknowledgments.}---It is a pleasure to thank N.~Arkani-Hamed and S.~Gubser for useful discussions. P.C. acknowledges the support of the IBM Einstein Fellowship. D.L.N., M.S., and G.T. are grateful to the IAS for hospitality during their work on this project. M.Z. is supported in part by  NSF Grants No. PHY-0855425, and No. AST-0907969, PHY-1213563 and by the David and Lucile Packard Foundation.

\end{document}